\documentclass[twocolumn,preprintnumbers,amsmath,amssymb]{revtex4}%

\usepackage{graphicx}
\usepackage{subfigure}
\usepackage{dcolumn}
\usepackage{bm}

\begin{document}

\title{\bf Spontaneous Formation of Dipolar Metal Nanoclusters}
\author{Elizabeth A. Sokol, Sara E. Mason, Valentino R. Cooper and Andrew M. Rappe}
\affiliation{  The Makineni Theoretical Laboratories, Department of
Chemistry\\ University of Pennsylvania, Philadelphia, PA 19104--6323 }%

\date{\today}

\begin{abstract}
The adsorption of Ag$_3$ and Ag$_4$ clusters on the
$\alpha$-Al$_2$O$_3$(0001) surface is explored with density functional
theory.  Within each adsorbed cluster, two different cluster-surface
interactions are present.  We find that silver clusters simultaneously
form both ionic bonds with surface oxygen and intermetallic bonds with
surface aluminum.  The simultaneous formation of disparate electronic
structure motifs within a single metal nanoparticle is termed a
"dipolar nanocluster".  This coexistence is ascribed to the similar
bond enthalpies of Ag--Al and Ag--O bonds, and its importance for
nanoparticle catalysis is highlighted.
\end{abstract}
\maketitle

There is great fundamental interest in understanding how transition
metals and oxides are affected by contact with each other.  Since
transition metal/oxide interfaces exert significant controllable
influence on material properties, many current and potential
applications rely on these heterostructures, including catalysts for
automotive pollution control~\cite{Gates95p511} and fuel
cells~\cite{Breen02p65}, as well as nanoscale
biosensors.~\cite{Whitney05p20522} Recent
theoretical~\cite{Eichler03p205408,Asthagiri06p125432,Finnis96p5811}
and experimental~\cite{Sanchez99p9573,Lee04p5682,Benz05p081102}
studies of size-selected nanocluster deposition onto oxides highlight
size dependence of the chemical and physical properties.  Furthermore,
numerous experimental and theoretical investigations emphasize the
role of the oxide support in changing the catalytic ability of these
metal-oxide
systems.~\cite{Roberts90p5337,Petrie94p8098,Haruta97p153,Walter01p44,Bozo01p393,Molina03p206102,Chen04p252,She06p79}

The combination of Ag and Al$_{2}$O$_{3}$ is special, because of the
close competition between ionic (Ag--O) and intermetallic (Ag--Al)
bonding.  Ag and Au have similarly low oxide formation
energies.~\cite{Feng05p115423} However, Au bonds to Al much more
strongly than to O, whereas the bonds that Ag makes to Al and to O
(Ag--Al$\approx$1.91~eV and Ag--O$\approx$2.28~eV) are more similar
than any other element (only Cu is close).~\cite{CRCHandbook}

In the present Letter, we report a novel consequence of this bonding
competition: small supported Ag$_n$ clusters exhibit two coexisting
structural and electronic relationships to an
$\alpha$-Al$_2$O$_3$(0001) ($\alpha$-alumina) substrate.  The
proximity of such different states within one cluster is fundamentally
interesting, and has ramifications for understanding and improving
noble metal nanocluster catalysis.  We use first-principles density
functional theory (DFT) to study the bonding of three- and four-atom
Ag clusters to the Al-terminated (0001) surface of
$\alpha$-Al$_2$O$_3$.  We find that bonding competition strongly
influences the stable cluster adsorption geometries, and is directly
responsible for inducing unusual electronic states in these clusters.

DFT calculations were performed with a generalized-gradient
approximation exchange-correlation functional.~\cite{Perdew96p3865}
Geometry optimizations were carried out using an in-house code, and
calculation of Born effective charges~\cite{Kingsmith93p1651} and
orbital-projected density of states (PDOS) were done using the ABINIT
software package.~\cite{Gonze92p3603} All calculations were converged
using a $2\times2\times1$ grid of Monkhorst-Pack
$k$-points.~\cite{Monkhorst76p5188} Norm-conserving optimized
pseudopotentials~\cite{Rappe90p1227} with the designed nonlocal method
for metals~\cite{Ramer99p12471,Grinberg01p201102} were constructed
using the OPIUM pseudopotential package.~\cite{Opium} The Kohn-Sham
orbitals are expanded in a plane-wave basis set truncated at 50~Ry.

The $\alpha$-Al$_2$O$_3$ surface is modeled by a slab geometry
supercell with an in-plane $(\sqrt{3}\times\sqrt{3})R30^\circ$ unit
cell and periodic boundary conditions.  The slabs consist of five
Al$_{3}$O$_{9}$Al$_{3}$ tri-layers, making the surfaces Al-terminated.
At least 12~\AA\ of vacuum separate periodic images in the (0001)
direction.  The theoretical Al$_2$O$_3$ in-plane lattice constant of
4.798~\AA\ was used (4.759~\AA\ experimental~\cite{Lee85p247}).

At equilibrium, the surface Al layer is only $z\approx 0.1$~\AA\ above
the O layer. In-plane relaxation yields nearest-neighbor O--O
distances from 2.63~\AA\ to 2.94~\AA.  Structural details of the
alumina surface are consistent with other modeling
studies~\cite{Ruberto03p195412}.

Chemisorption of Ag on the $\alpha$-Al$_2$O$_3$ (0001) surface
strongly favors cluster
formation~\cite{Zhukovskii98p73,Zhukovskii02p343} because of 4--5\%
mismatch between the surface O--O distance (2.76~\AA) and the Ag--Ag
distance (2.88~\AA) in bulk Ag.  To find the minimum-energy structures
for Ag$_3$, for planar Ag$_4$, and for pyramidal Ag$_4$, each cluster
was initially placed three different ways. Interfacial Ag atoms were
started at top, bridge, or hollow sites of the oxygen lattice, and
each system was allowed to relax fully.  In each calculation, the top
two surface tri-layers of alumina were allowed to relax in all
directions, the third tri-layer was relaxed in the direction normal to
the surface plane, and the bottom two tri-layers were constrained to
their bulk alumina coordinates.

\begin{figure}
  \centering \subfigure[]{
    \label{fig:ChdTop}
    \includegraphics[width=1.5in]{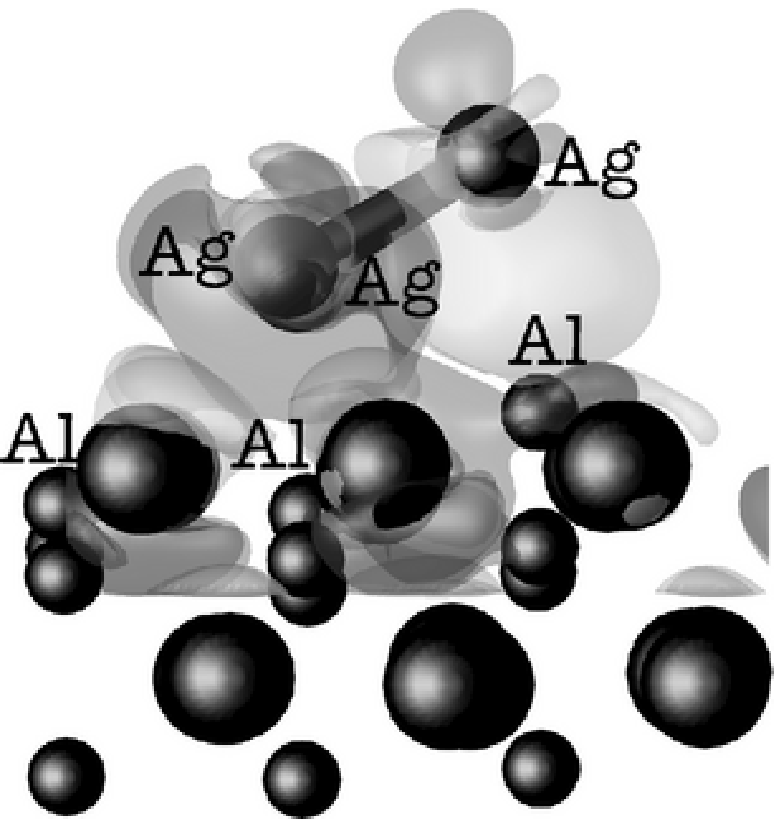}}
  \subfigure[]{
    \label{fig:ChdSide}
    \includegraphics[width=1.3in]{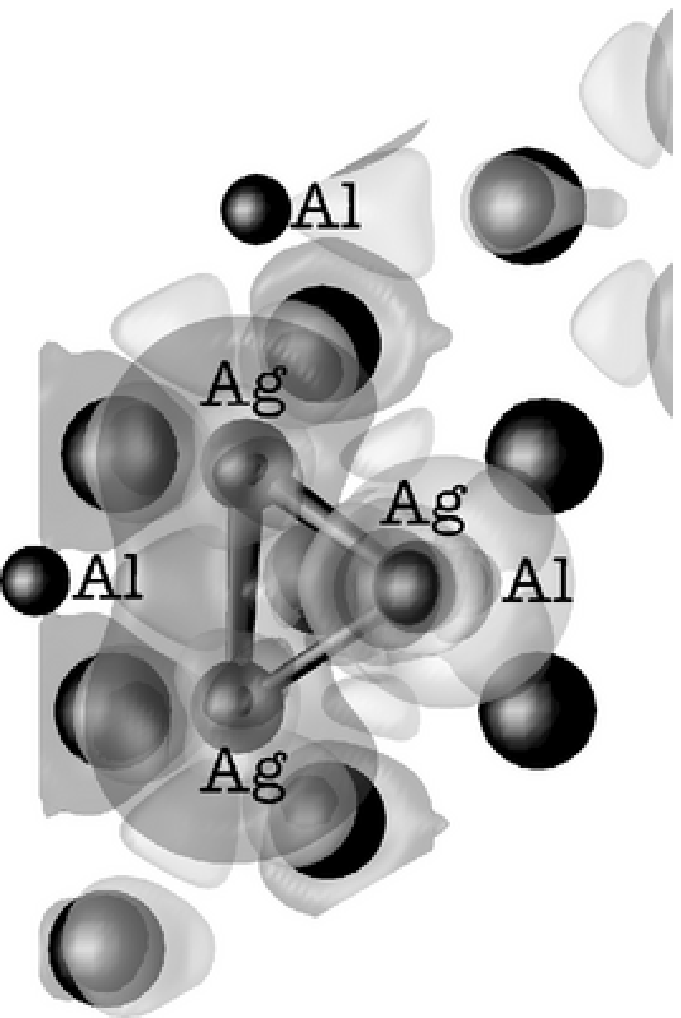}}
     \caption{Induced charge density $\Delta\rho$ diagrams for the
       optimized structure of Ag$_3$/Al$_2$O$_3$.  O, Ag (connected),
       and Al are shown as black spheres of decreasing size.
       Adsorption causes electron flow from dark to light regions.
       Topmost Ag and Al atoms are labeled.  Iso-surface values are
       $\pm$0.02~$e^-$/\AA$^3$.  (a) Side view. (b) Top view.}
     \label{fig:TrimerChD}
\end{figure}

For all starting positions, geometry optimization of
Ag$_3$/Al$_2$O$_3$ leads to the same final structure
(Figure~\ref{fig:TrimerChD}), with all Ag atoms in hollow sites of the
top O layer lattice, one above a surface Al atom, and two above
subsurface Al atoms.  The cluster chemisorption energy $E_{\rm ads}$
is 2.09~eV.  The cluster tilts by 30$^\circ$ with respect to the
surface plane, with the Ag above the surface Al farthest from the
surface.  Defining {\em z} as the distance of Al atoms from the
topmost oxygen layer in the surface normal direction, the values in
bare Al$_{2}$O$_{3}$ are 0.096~\AA\ for the topmost Al of the unit
cell.  In response to cluster adsorption, the nearest surface Al atoms
move vertically by $\Delta z=$-0.47, -0.55, and +0.57~\AA.  One Ag
and one Al atom shift upward together.  The inward relaxation of the
the other nearby surface Al atoms induced by metal cluster adsorption
has been observed in other theoretical studies~\cite{Gomes03p6411}.

The relaxed interatomic distances strongly suggest two different
bonding motifs for the Ag$_3$ cluster.  The raised Ag and Al make a
short 2.65~\AA\ bond, which is just about the sum of their covalent
radii ($r_{\rm Ag}$=1.53~\AA, $r_{\rm Al}$=1.18~\AA).  The shortest
distance from this Ag to O is 3.66~\AA, much longer than the sum of
their ionic radii ($r_{\rm Ag^{+}}$=1.14~\AA, $r_{\rm
O^{2-}}$=1.24~\AA).  The other two Ag atoms have short
$\approx$2.46~\AA\ distances to O, quite in line with ionic bonding.
These Ag atoms also have long distances of 2.95~\AA\ to subsurface Al,
suggesting little if any covalent interaction in that case.  So the
bond length data can be summarized as a covalent
(intermetallic, IM) bond between the raised Ag and Al, and ionic Ag--O
bonds for the other two Ag atoms.

We study $\Delta\rho$, the change in charge density induced by the
cluster adsorption, to visualize the electronic cluster-surface
interactions.  The side and top views of the iso-surfaces for
$\Delta\rho$ are shown in Figure~\ref{fig:TrimerChD}, with electronic
charge flowing from dark to light regions.  Figure~\ref{fig:TrimerChD}
shows gain of charge between the raised Ag and Al, indicative of
Ag--Al IM bond formation.  The other two Ag atoms show a significant
loss of electrons, with a corresponding gain of electrons for the
nearest topmost surface oxygens.  Therefore, these interactions are
chiefly ionic bonds.

This electronic description of Ag$_3$ adsorption also provides insight
into the observed surface relaxation.  The ionic bonds formed between
Ag and O leave these O atoms less capable of bonding to surface Al
atoms, using a bond-valence argument.~\cite{Brown81p1} Therefore,
these adjacent surface Al atoms relax inward (below the top O layer),
so they can form bonds with subsurface oxygen.  The increase
in electron density near the IM Al reduces its electrostatic
interaction with oxygen and results in the observed outward
relaxation.

We propose referring to these clusters as ``(electric) dipolar
nanoparticles.''  (Magnetic) dipolar nanoparticles have been
reported~\cite{Ditsch05p6006}, but we know of no previous report showing
spontaneous formation of electric dipoles on nano-sized supported
metal particles.

To make precise the magnitude and orientation of the nanoparticle
dipole, we apply the modern theory of
polarization.~\cite{Kingsmith93p1651} The Born effective charge tensor
$Z^*$ is found~\cite{Gonze97p10355,Ghosez98p6224}, where the tensor
element $Z^*_{\alpha ij}$ gives the change in electric polarization
component $P_i$ as atom $\alpha$ moves along direction $j$.  There is
no requirement that this mixed second derivative tensor be symmetric,
and the tensors of the supported clusters are highly anisotropic.  We
have obtained the principal values of the charge tensors and for each
supported Ag atom the largest principal values (of each sign, where
applicable) and their corresponding principal directions are reported
in Table~\ref{table:Zstar}.

\begin{table}
\caption{Principal values of Born effective charge tensors and
electronic {\em d}-band centers $\epsilon_{d}$ for supported cluster
Ag atoms.  For each atom, the largest principal value of each sign,
along with the corresponding principal direction, are reported.  The
principal directions (shown in Figure~\ref{fig:zstarfig}) in all cases point inward toward the cluster center and along the surface normal.  The angle $\theta$ between the principal direction and the surface normal are tabulated.}
\begin{tabular}{lccc}
&\multicolumn{1}{c}{Principal Value}
&\multicolumn{1}{c}{$\theta$,$^\circ$}
&\multicolumn{1}{c}{$\epsilon_{d}$, eV}\\
\hline
Ag$_{3}$ Ionic   &  1.11 & 32 & -3.42\\
Ag$_{3}$ IM      & -0.23 & 60 & -4.32\\
Boat ionic       &  1.45 & 31 & -3.14\\
Boat IM          & -0.42 & 41 & -3.66\\
Boat IM          &  0.11 & 87 & -3.66\\
Candlestick ionic &  1.22 & 26 & -3.63\\
Candlestick IM    & -0.35 & 59 & -4.56\\
Pyr ionic         & 1.48  & 23  & -3.39 \\
Pyr IM            &-0.27  & 0  & -4.16\\
Pyr IM            & 0.36  & 85  & -4.16\\
Top Ag            & -0.11 & isotropic & -3.78\\
\end{tabular}
\label{table:Zstar}
\end{table}

\begin{figure}
  \centering
  \subfigure[]{
    \includegraphics[width=1.0in]{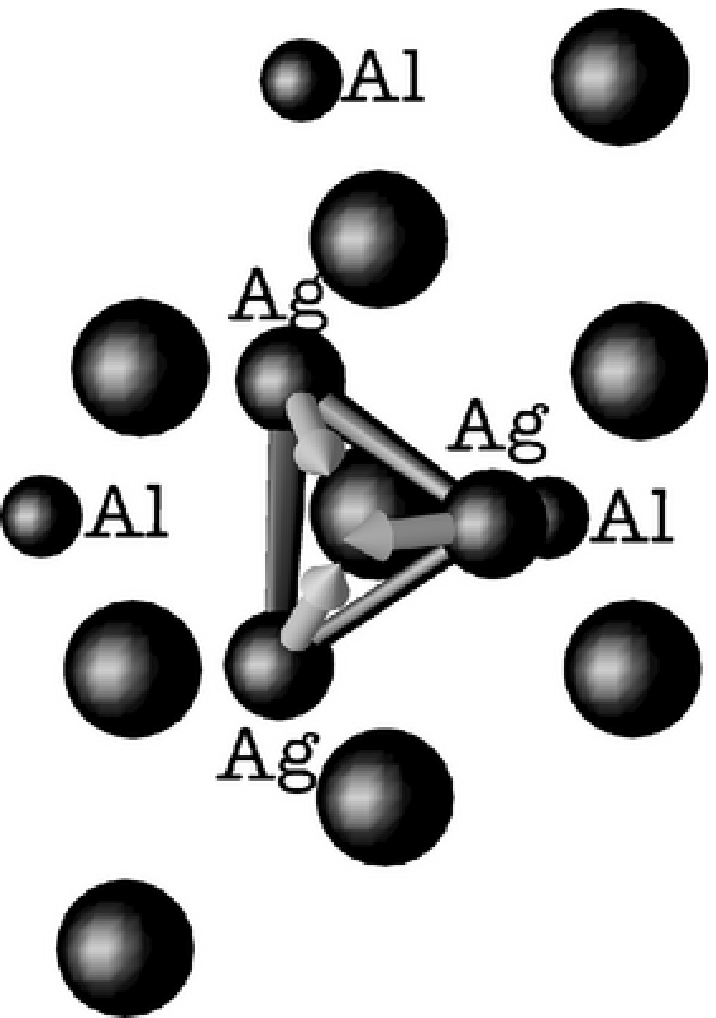}}
  \hspace{0.2in}
  \subfigure[]{
    \includegraphics[width=1.0in]{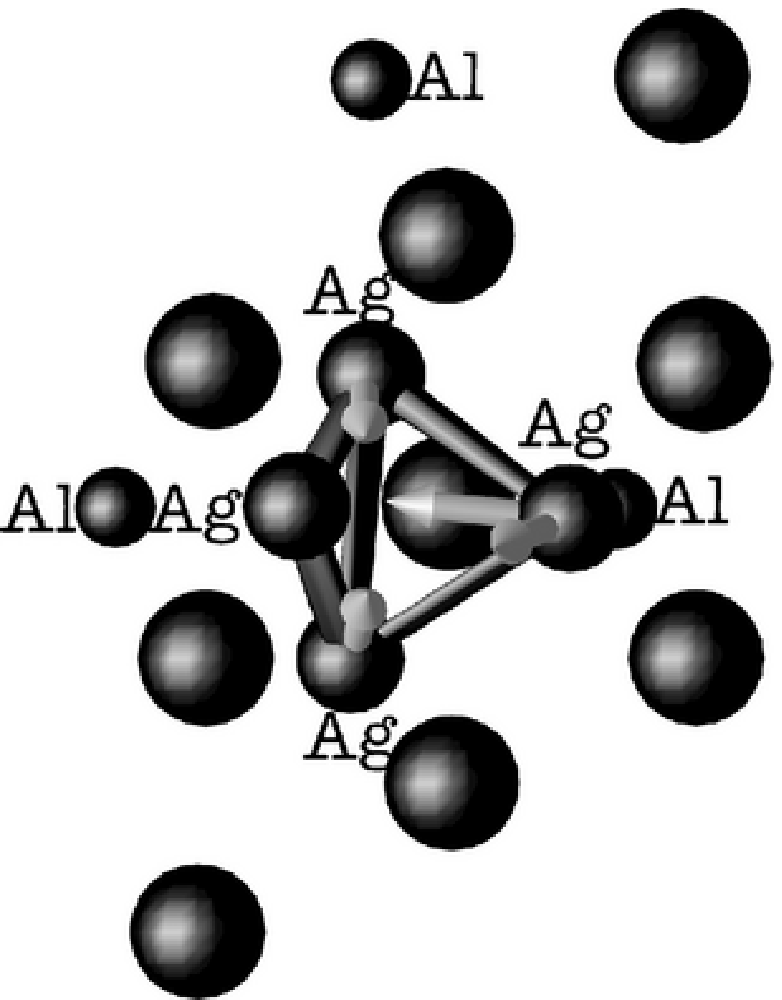}}
  \subfigure[]{
    \includegraphics[width=1.0in]{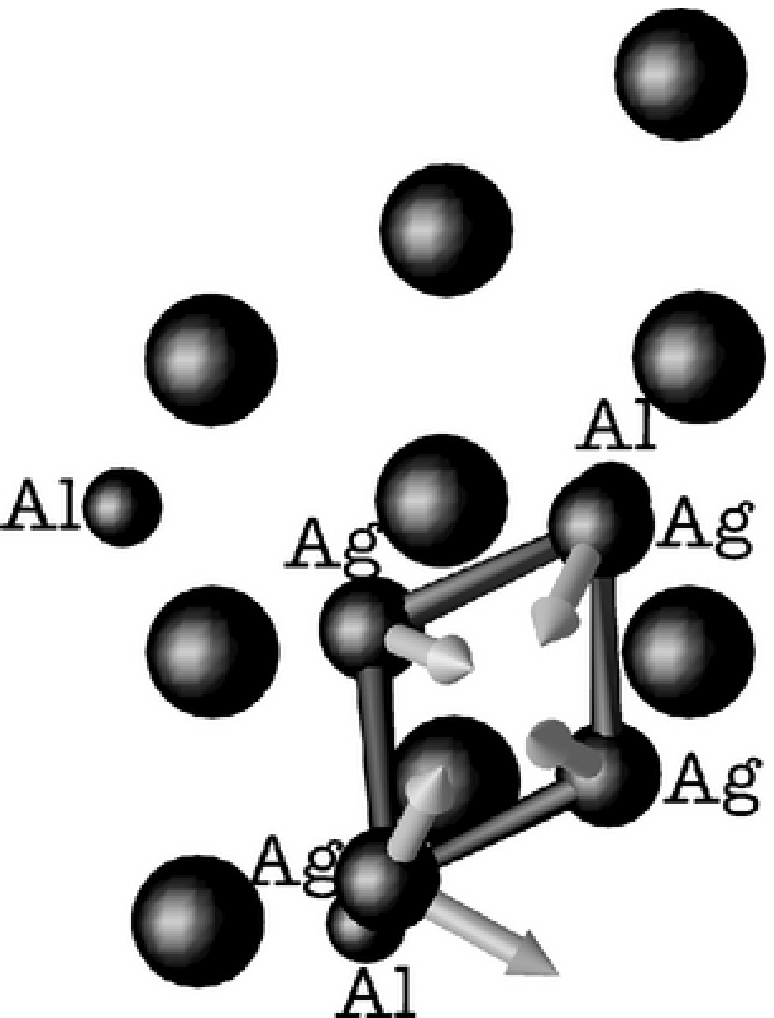}}
  \hspace{0.2in}
  \subfigure[]{
    \includegraphics[width=1.0in]{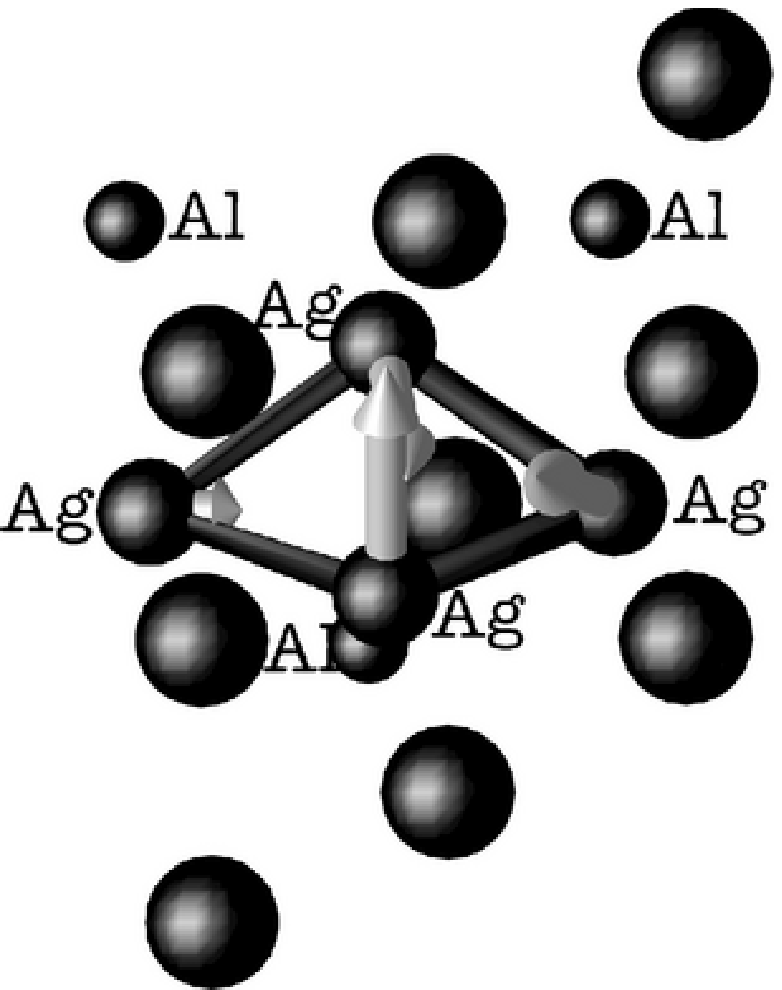}}\\
     \caption{Top views of optimized cluster geometries.  O, Ag
       (connected), and Al are shown as black spheres of decreasing
       size.  Unit vectors of the principal directions of the Born
       effective charge principal values are indicated by arrows.
       Only the topmost nine O and three Al surface atoms of the unit
       cell are shown.  (a) Ag$_{3}$ (b) Pyr (c) Boat (d) Candlestick}
     \label{fig:zstarfig}
\end{figure}

The dynamical charge values support the above interpretation of the
Ag/Al$_2$O$_3$ interactions: ionically bound Ag atoms lose charge
through interaction with surface O, resulting in positive principal
values.  The IM Ag atom gains charge from the bonding with surface Al,
resulting in a modest negative $Z^*$, but most of the charge is
shared, not closely associated with Ag motion.  In some cases (boat,
pyr) the IM atoms have a more complicated Born effective charge
tensor, with positive and negative principal values.  The principal
directions all point inward to the center of the cluster and out of
the surface plane (Figure~\ref{fig:zstarfig}).  Increased negative
charge on the IM Al atom and ionic O atoms in the adsorbed geometry
relative to the bare surface compensate for the net positive charge
localized on the cluster.

\begin{figure}
\includegraphics[width=3.0in]{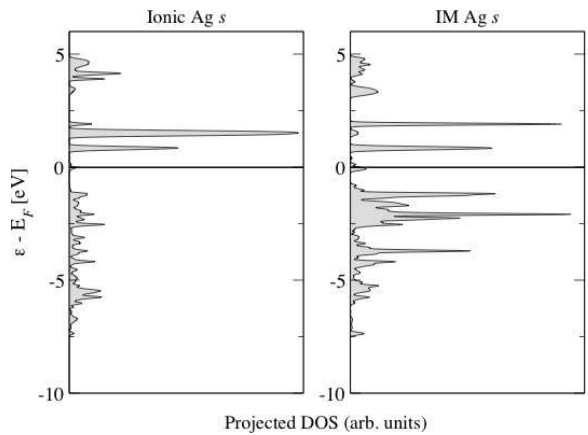}
\caption{{Density of states projected onto the {\em s}-orbital of Ag
atoms in Ag$_{3}$/Al$_{2}$O$_{3}$.}}
\label{fig:Agdos}
\end{figure}

Ionic and IM Ag-surface interactions are also revealed in PDOS
analysis.  Covalent IM bonding causes orbitals to mix, leading to
intensity in the IM Ag {\em s}-PDOS below the Fermi level
(Figure~\ref{fig:Agdos}).  The {\em s} orbital of the ionic Ag shows
less intensity at bonding levels, and is dominated by substantial
intensity above the Fermi level.  The fractional fillings of the IM
and ionic Ag {\em s} orbitals are 0.69 and 0.35, respectively.

The projection onto Ag {\em d}-states furhter demonstrates how the
coexistence of two cluster/surface interactions affects the reactivity
of the supported Ag atoms.  Our analysis shows that the filling of the
supported Ag {\em d}-bands is constant and near unity for all Ag
atoms.  Table~\ref{table:Zstar} lists the average energy of the {\em
d}-band projections ($\epsilon_{d}$) of supported Ag atoms with
respect to the Fermi level.  $\epsilon_{d}$ is well established as a
predictive parameter for assessing reactivity~\cite{Hammer96p2141}.
The 0.90~eV shift in $\epsilon_{d}$ between the ionic and IM Ag atoms
in Ag$_{3}$/Al$_{2}$O$_{3}$ is larger then what can be achieved
through perturbations such as strain~\cite{Mavrikakis98p2819} and is
more in line with the extent of shift brought about by significant
coordination change~\cite{Hammer97p31} or introduction of a metal
hetero metal atom to a surface~\cite{Ruban97p421}.

Having built a model of the chemisorption bonding in the
Ag$_3$/Al$_2$O$_3$ system, we examine how the chemisorption of Ag$_4$
onto Al$_2$O$_3$ expands these ideas.  As with the
Ag$_{3}$/Al$_{2}$O$_{3}$ system, calculations of $Z^*$ and
PDOS were carried out on all optimized cluster geometries, and key
results are discussed.

Pyramidal Ag$_4$ (``pyr'') bonds to Al$_{2}$O$_{3}$ ($E_{\rm
ads}$=1.96~eV) such that the three base atoms tilt and interact with
the surface very similarly to Ag$_3$/Al$_2$O$_3$. This and near-zero
principal values of $Z^*$ for the pyramidal top Ag atom imply that the
base screens the top atom from electrostatic interaction with the
surface, consistent with previous results concerning the length scale
of metal-oxide interactions.~\cite{Cooper05p081409R} A single Ag--Al
IM bond is formed, and the other two Ag atoms participate in ionic
interactions with the topmost surface oxygen.  The Al $\Delta z$
values are similar to those found in the Ag$_3$/Al$_2$O$_3$ structure.
The top Ag atom of the pyramid in the supported geometry is not in the
Ag trimer hollow, but is shifted almost to the bridge site, above the
two charge-depleted ionic Ag atoms.

The bonding competition between Ag--Al and Ag--O directly leads to two
metastable minima for the planar Ag$_4$/Al$_2$O$_3$ system.  Each
starting structure is a parallelogram parallel to the surface.  The
cluster buckles significantly as it chemisorbs.  We find two local minima,
each with all Ag atoms in hollow sites of the O lattice; the
structures differ in their registry with the top Al sublattice.  The
ground state (``boat'') has two Ag atoms above surface Al, while the
other local minimum (``candlestick'') has only one Ag--Al interaction.

The boat and candlestick Ag$_4$ clusters have $E_{\rm ads}$ values of
2.14~eV and 1.49~eV, respectively.  The boat is energetically favored,
suggesting that a balance of IM and ionic bonding stabilized both.  In
fact, the boat exhibits larger positive and negative $Z^{*}$ values
than the candlestick.  Therefore, the energetically favored Ag$_{4}$
boat cluster will exhibit even stronger dipolar nanoparticle
properties than Ag$_{3}$.

In conclusion, we find that bonding competition between Ag--Al and
Ag--O gives rise to ionic Ag--O and intermetallic Ag--Al interactions
between Ag cluster atoms and the alumina surface.  The proximal
coexistence of these interactions results in the formation of dipolar
nanoparticles.  The electronic and structural effects are closely
related, with IM and ionic Ag-surface bonding favoring outward and
inward Al motion, respectively.  We find consistent results and
interpretations of induced charge density, Born effective charges, and
projected density of states in all four optimized cluster geometries
(Ag$_{3}$, pyr, boat, and candlestick).  Principal values of $Z^*$
show that ionic Ag atoms with positive charge and IM atoms with
negative charge can be clearly distinguished, while stacked Ag atoms
are mostly screened.  The coexistence of ionic and IM bonding in
supported clusters may be possible to create in other metal-oxide
combinations, where the oxide and IM bond enthalpies are inherently
similar or by tuning the competition by means of surface modification.

This work was supported by the Air Force Office of Scientific
Research, Air Force Materiel Command, USAF, under Grant
No. FA9550-04-1-0077.  Computational support was provided by the
Defense University Research Instrumentation Program, and the
High-Performance Computing Modernization Program of the
U. S. Department of Defense.

\end{document}